\documentstyle[12pt]{article}
\font\Sets=msbm10
\def\Integer {\hbox{\Sets Z}}    
\def\Complex {\hbox{\Sets C}}
\def\be{\begin{equation}}       \def\ba{\begin{array}}
\def\ee{\end{equation}}         \def\ea{\end{array}}
\def\bea {\begin{eqnarray}}      \def\eea {\end{eqnarray}}
\def\bean{\begin{eqnarray*}}    \def\eean{\end{eqnarray*}}
  \def\ti  {\widetilde}    \def\dag {\dagger}
\def\la  {\lambda}   \def\eps{\varepsilon}    \def\ph{\varphi}

\def\im    {\mathop{\rm Im}   \nolimits}

\def\ker  {\mathop{\rm Ker} \nolimits}
\def\RA {\ \Rightarrow\ }         \def\LRA {\ \Leftrightarrow\ }
\def\qed   {\vrule height0.6em width0.3em depth0pt}
\def\defeq {\stackrel{\mbox{\rm\small def}}{=}}
\newtheorem{exi}{Example}

\begin{document}
\title{Elementary Darboux transformations and factorization}

\author{ F.Musso$^\sharp$ and A.Shabat$^\flat$}
\maketitle
\begin{center} $^\sharp$ Dipartimento di Fisica E. Amaldi, Universit\`{a}
degli Studi di Roma Tre\\
and\\
Istituto Nazionale di Fisica Nucleare, Sezione di Roma Tre\\
Via della Vasca Navale 84, 00146, Rome, Italy
\footnote{E-mail:musso@fis.uniroma3.it}
\end{center}
\begin{center} $^\flat$Landau Institute for Theoretical Physics, Chernogolovka,
Russia\\
and\\
Dipartimento di Fisica E. Amaldi, Universit\`{a}
degli Studi di Roma Tre\\
Via della Vasca Navale 84, 00146, Rome, Italy
\footnote{E-mail:shabat@fis.uniroma3.it}
\end{center}
\thispagestyle{empty}
\begin{abstract} A general theorem on factorization of matrices with polynomial
 entries is proven and it is used to reduce polynomial Darboux
matrices to linear ones. Some new examples of linear Darboux
matrices are discussed.
 \end{abstract}
\thispagestyle{empty}

\section{Introduction}
In the continuous case, for spectral problems with  $m\times
m$ matrix potentials $U$ polynomial in $\la$
\be\label{SP}  {d\over
dx}\Psi= U(x,\la)\Psi  \ee
 the construction of Darboux transformation
\be\label{hatpsi} \Psi\to\hat\Psi=Q(x,\la)\Psi,\quad
\frac{d}{dx}Q=\hat{U}Q-QU\RA {d\over dx}\hat \Psi= \hat U(x,\la)\hat \Psi
\ee
  can be based on the {\it elementary} Darboux transformations, for which
\be \label{sha}
Q(x,\la)=(\la-\alpha)P(x)+(\la-\beta)(E-P(x)),\quad
P^2=P.
\ee
Here, $E$ denotes the identity matrix and the projector $P$ is uniquely 
defined by the null-spaces of $Q:$
\[ \ker Q|_{\la=\alpha}=\im P=M ,\quad \ker Q|_{\la=\beta}=\ker P=N \]
If $v(x) \in M$ and $w(x) \in N$, then the following equations have to be satisfied:  
\be\label{mnx}
 \frac{d}{dx}v(x)=U(x,\alpha)v(x),\quad
 \frac{d}{dx}w(x)=U(x,\beta)w(x).
\ee
 The zeroes of the determinant
 define the eigenvalues $\la=\alpha,\, \beta$ introduced by the Darboux
 transformation (\ref{sha}). In particular, for the Zakharov-Shabat spectral
 problems (see, e.g., \cite{ZS2}),
 \be\label{ZS} \hat{U}=\la\sigma+q(x) ,\quad q(x)=[\sigma, \ti{q}(x)],    \ee
 where $\sigma$ is
 a constant matrix with simple eigenvalues, the function
 $\hat\Psi$ satisfies the equation (\ref{SP}) with the potential
\[                       U=\hat U+[Q,\sigma].        \]
 The case
 \be\label{*} \beta=\bar\alpha,\quad P^*=P  \ee
 where $*$ means hermitian conjugation, corresponds to the self-adjoint
 potentials
 \[   U^*(x,\bar\la)=U(\la,x),\quad q^*=q.   \]

 Main Theorem in Section 2 \footnote{An analog of this theorem was recently 
used by \cite{Do} in
 order to construct an approximate solution of Riemann-Hilbert problem. 
See also \cite{Ha}.} formulates the necessary and sufficient conditions
 on the $N$th degree matrix polynomial $A(\la)$ to be represented as
 the product of $N$ first order polynomials (\ref{sha}), (\ref{*}).
 
 In application to Darboux transformations this factorization $A=Q_N\cdots
 Q_1$ gives rise the sequence of spectral problems (\ref{SP})
\be\label{Psi}
 {d\over dx}\Psi_j= U_j(x,\la)\Psi_j,\quad
 \Psi_{j+1}= Q_j(x,\la)\Psi_j,
  \ee
related with each other by the transformations (\ref{hatpsi}),(\ref{sha}). 
Observe that the composition of Darboux transformations
$\Psi_{j+1}= B_j\Psi_j$ is also a Darboux transformation since
(Cf. (\ref{hatpsi}))
\be\label{bub}
 B_{j,x}=U_{j+1}B_j-B_jU_j,\quad B\defeq B_n\cdots B_1\RA B_x=U_{n+1}B-BU_{1}.
\ee

For a scalar differential operator $A$, the {\bf elementary} Darboux
transformation $\Psi\to\hat\Psi$ is defined by the first order differential
operator:
\[ A\Psi=\la\Psi,\quad \hat\Psi=(D-f)\Psi,\quad \hat A\hat\Psi=\la\hat\Psi \quad \hat A=(D-f)A (D-f)^{-1}.\]
For the operator $A$ of order $N$, an analog of Main Theorem yields
the relation \be\label{dar}
A\Psi=\frac{<\ph_1,\dots,\ph_N,\Psi>}{<\ph_1,\dots,\ph_N>}=
(D-f_{N})\cdots(D-f_2)(D-f_1)\Psi, \ee where $\ph_j\in\ker A$ and
\[
  <\ph_1,\dots,\ph_l>\defeq \det(D^{k-1}(\ph_j)), \quad j,k=1,\dots,l,\quad
  D=\frac{d}{dx}.
 \]
 The proof of (\ref{dar}) is obtained by induction (Cf. Main Theorem)
 based upon the identity
 valid for any smooth functions $\ph_j:$
 \[ <\ph_1,\dots,\ph_m>=\ph_1<\hat\ph_2,\dots,\hat\ph_m>,\quad
 \hat\ph_j=(D-f_1)\ph_j,\quad f_1=D\log\ph_1.
\]

In eq. (\ref{dar}) the factors $D-f_j$ are irreducible
(they are first order differential operators). Such property
appears to be crucial in the theory of dressing chains
\cite{VeSh}, \cite{99}, \cite{AdMaSh}, where the functions $f_j$
are used as basic dynamical variables. On the other hand, the Darboux
transformation (\ref{sha}) can be factorized into the composition
of two more elementary ones with null spaces $\ker P$ and $\im P,$
respectively. This factorization reformulated in terms of the
corresponding matrices $B_1$ and $B_2$ gives rise to the following
equations:
\[ \Psi(m+1,n)=B_1(m,n)\Psi(m,n),\quad \Psi(m,n+1)=B_2(m,n)\Psi(m,n),\]
\[ \Psi(m+1,n+1)=Q(m,n)\Psi(m,n),  \quad Q=B_1\circ B_2,  \]
where by definition \be\label{qmn} Q=B_1\circ B_2\LRA
Q(m,n)=B_1(m,n+1)B_2(m,n)=B_2(m+1,n)B_1(m,n). \ee Equations
(\ref{qmn}) show that, unlike the continuous case (\ref{bub}),
the Darboux transformations and the original discrete spectral problem
have the same nature and it is possible to replace one with the
other.

Thus, the problem of factorization of (\ref{sha}), considered in
Section 3, brings up the discrete spectral problems with
potentials linear in the spectral parameter $\la$. In the examples, we will
restrict ourself to the simplest $2\times2$ case but without the
assumption that $Q$ is factorized through {\em orthogonal} projectors 
(see (\ref{sha})).
In our considerations, the continuous independent variable $x$ 
will play the role of a parameter .

In the last subsection we consider solutions of the Yang-Baxter
equation linear in the spectral parameter $\la$.

\section{Main Theorem}
In order to prove that the Darboux transformations (\ref{sha}) are generic
ones, it sufficies to prove that the product of $N$ matrices (\ref{sha})
yields an $N$-th degree polynomial in $\la$ of a general form. We
will use the characteristic property of the polynomials
(\ref{sha}) as follows \be\label{qab} Q_{\alpha\beta}\defeq
(\la-\alpha)P+(\la-\beta)(E-P)\RA
Q_{\alpha\beta}Q_{\beta\alpha}=(\la-\alpha)(\la-\beta)E. \ee 
In case (\ref{*}), we have
\[   Q(\bar{\lambda})^*Q(\lambda)=(\la-\alpha)(\la-\bar\alpha)E \],
and such property holds, obviously, for the product of matrices
(\ref{sha}) with the orthogonal projectors $P^*=P.$

\paragraph{Theorem.}
Let $A(\lambda)$ be an $m \times m$ matrix with polynomial entries
in $\lambda$ satisfying the conditions:
\begin{eqnarray}
&& A(\bar{\lambda})^*A(\lambda)=\prod_{j=1}^N
(\lambda-\alpha_j)(\lambda-\bar{\alpha}_j) E,
\label{condition}\\
&& {\rm{lim}}_{\lambda \rightarrow \infty}
\frac{A(\lambda)}{\lambda^N}=E. \label{condition2}
\end{eqnarray} Then
$A(\lambda)$ can be factorized as
\begin{equation}
A(\lambda)= \prod_{j=1}^N \left((\lambda-\alpha_j)P_{\alpha_j}+
(\lambda-\bar{\alpha}_j)(E-P_{\alpha_j}) \right),
\label{factor}
\end{equation}
where the $P_{\alpha_j}$ are Hermitian projectors.

\noindent{\bf Proof:} We proceed by induction on $N$. If $N=0$,
then from (\ref{condition2}) and the Liouville theorem it follows that
$A(\lambda)=E$.

Now we show that if the statement of our theorem holds for $N-1$, it holds for $N$
as well. From (\ref{condition}) and (\ref{condition2}) it follows:
\begin{equation}
{\rm{det}}(A(\lambda))=\prod_{j=1}^N
(\lambda-\alpha_j)^{k_j}(\lambda-\bar{\alpha}_j)^{m-k_j}.
\label{Al}
\end{equation}
We select a zero $\alpha_j$ of ${\rm{det}}(A(\lambda))$; from
(\ref{Al}) it follows:
\begin{equation}
\begin{array}{l}
{\rm{dim(Ker}}(A(\alpha_j)))=k_j, \\
{\rm{dim(Ker}}(A(\bar{\alpha}_j)))=m-k_j.
\end{array} \label{dim}
\end{equation}
Let $|v_i \rangle_{i=1}^{k_j}$ and $|w_i \rangle_{i=1}^{N-k_j}$ be
an orthonormal basis for ${\rm{Ker}}(A(\alpha_j))$ and
${\rm{Ker}}(A(\bar{\alpha}_j))$, respectively. We denote by
$P_{\alpha_j}$ and $P_{\bar{\alpha}_j}$ the Hermitian projectors
\begin{eqnarray}
&& P_{\alpha_j} \equiv \sum_{i=1}^{k_j} |v_i \rangle \langle v_i|, \\
&& P_{\bar{\alpha}_j} \equiv \sum_{i=1}^{m-k_j} |w_i \rangle
\langle w_i|.
\end{eqnarray} Let us prove that
\begin{equation}
P_{\bar{\alpha}_j}=(E-P_{\alpha_j}). \label{l1}\\
\end{equation}
To prove (\ref{l1}) is equivalent to proving that
$\Complex^m$ admits an orthogonal decomposition
\begin{equation}
\Complex^m= {\rm{Ker}}(A(\alpha_j)) \oplus
{\rm{Ker}}(A(\bar{\alpha}_j)). \label{ortho}
\end{equation}
From (\ref{condition}) it follows:
\begin{equation}
 A(\alpha_j) A(\bar{\alpha}_j)^*=0,
\end{equation}
that is
\begin{equation}
{\rm{Im}}(A(\bar{\alpha}_j)^*) \subset {\rm{Ker}}(A(\alpha_j));
\end{equation}
but from (\ref{dim}) we have:
\begin{eqnarray*}
&& {\rm{dim(Im}}(A(\bar{\alpha}_j)^*))=
m-{\rm{dim(Ker}}(A(\bar{\alpha}_j)))=k_j,\\
&& {\rm{dim(Ker}}(A(\alpha_j)))=k_j,
\end{eqnarray*}
whence it follows:
\begin{equation}
{\rm{Im}}(A(\bar{\alpha}_j)^*) = {\rm{Ker}}(A(\alpha_j)).
\label{fund}
\end{equation}
Equation (\ref{fund}) is equivalent to
\begin{equation}
{\rm{Ker}}(A(\bar{\alpha}_j)) = {\rm{Im}}(A(\alpha_j)^*)
\label{fund2}
\end{equation}
whence it follows
\begin{displaymath}
\Complex^m= {\rm{Ker}}(A(\alpha_j)) \oplus
{\rm{Im}}(A(\alpha_j)^*)= {\rm{Ker}}(A(\alpha_j)) \oplus
{\rm{Ker}}(A(\bar{\alpha}_j)).
\end{displaymath}
Now the matrix
\begin{equation}
A(\lambda) \left[ (\lambda-\bar{\alpha}_j) P_{\alpha_j}+
(\lambda-\alpha_j) (E-P_{\alpha_j}) \right]
\end{equation}
vanishes at $\lambda=\alpha_j$ and
$\lambda=\bar{\alpha}_j$. Consequently, we have:
\begin{equation}
A(\lambda) \left[ (\lambda-\bar{\alpha}_j) P_{\alpha_j}+
(\lambda-\alpha_j) (E-P_{\alpha_j})
\right]=(\lambda-\alpha_j)(\lambda-\bar{\alpha}_j)\hat{A}(\lambda)
\end{equation}
for a certain matrix $\hat{A}(\lambda)$. Then
\begin{equation}
A(\lambda)= \hat{A}(\lambda) \left[ (\lambda-\alpha_j)
P_{\alpha_j}+ (\lambda-\bar{\alpha}_j) (E-P_{\alpha_j})
\right]
\end{equation}
Since $P_{\alpha_j}$ is Hermitian, it follows that
$\hat{A}(\lambda)$ satisfies
\begin{equation}
\hat{A}(\bar{\lambda})^*\hat{A}(\lambda)=\prod_{k \neq j}^N
(\lambda-\alpha_k)(\lambda-\bar{\alpha}_k) E.
\end{equation}
Moreover,
\begin{equation}
{\rm{lim}}_{\lambda \rightarrow \infty}
\frac{A(\lambda)}{\lambda^N}={\rm{lim}}_{\lambda \rightarrow
\infty} \frac{\hat{A}(\lambda)}{\lambda^{N-1}}=E. \qquad \qed
\end{equation}

In the case of distinct zeroes $\alpha_j\ne \alpha_k$, the
orthogonal projectors $P_{\alpha_j}$ in (\ref{factor})
are uniquely defined by the subspaces \be\label{Mj} M_j\defeq\ker
A(\la)|_{\la=\alpha_j} \ee and an ordering of zeroes
$\alpha_1,\dots,\alpha_N.$ Indeed, for instance, for $N=2$, we have
\[
A(\la)=Q_1(\la)Q_2(\la),\quad Q_j(\la)=
(\la-\alpha_j)P_{\alpha_j}+(\la-\bar{\alpha}_j)(E-P_{\alpha_j}), \quad
j=1,\, 2;
\]
with $P_{\alpha_j}^*=P_{\alpha_j}$ and
\[
\im P_{\alpha_2}=\ker(E-P_{\alpha_2})=\ker Q_2(\alpha_2)=
\ker Q_1(\alpha_2)Q_2(\alpha_2)=M_2\],
since $\det Q_1(\alpha_2)\ne 0.$ Similarly
\be\label{imp1}
\im P_{\alpha_1}=Q_2(\alpha_1)M_1,\quad\mbox{since}\quad
M_1=\ker(E-P_{\alpha_j})Q_2(\alpha_1).
\ee

In order to construct a dressing chain, we have to factorize (see the next
section) the basic Darboux transformation (\ref{qab}) further:
\be\label{qbb}
 Q_{\alpha\beta}(\la)=(\la-\alpha)P_{\alpha\beta}+
 (\la-\beta)(E-P_{\alpha\beta})= B_2(\la)B_1(\la)
 \ee
 where these two factors $B_j(\la)=\sigma_j \la+b_j$ are linear in $\la$ and
 \[
\det Q_{\alpha\beta}(\la)=(\la-\alpha)^k(\la-\beta)^{m-k} \RA
\det B_1=(\la-\alpha)^k,\quad\det B_2=(\la-\beta)^{m-k}.
\]
Evidently,
\be\label{12} \ker B_1(\beta)=\im B_2(\alpha)=\ker
P_{\alpha\beta}, 
\ee 
since, due to (\ref{qbb}) and the fact that $\det B_1(\alpha)\ne0$, we have
\[
 Q_{\alpha\beta}(\alpha)= (\alpha-\beta)(E-P_{\alpha\beta})=
  B_2(\alpha)B_1(\alpha)\RA \im B_2(\alpha)=\im(E-P_{\alpha\beta})=
  \ker P_{\alpha\beta}.
\]

\section{$\mathbf{2 \times 2}$ case}
In this section we consider some interesting, from our point of
view, classes of $2\times 2$ matrices linear in $\la$ related with
(\ref{sha}). First, let us consider the chain of matrices
$B_j,\, j \in \Integer:$
 \be\label{odd} B_j=\left(\ba{cc}
 \la+h_j & \frac12 u_j\\
 \frac12v_{j+1} & \hfill 1
\ea\right), \quad h_j= \frac14 u_jv_{j+1}-\beta_j \quad\mbox{for}\quad j \quad \mbox{odd} , \ee
\be\label{even} B_j=\left(\ba{cc}
 1 & -\frac12 u_{j+1}\\
 -\frac12v_j & \la+h_j
\ea\right),\quad h_j= \frac14 v_ju_{j+1}-\beta_j
\quad\mbox{for}\quad j \quad\mbox{even}. \ee
Thus
 \[  \det B_j=\la-\beta_j \quad\mbox{for all}\quad j \]
 and (Cf (\ref{12})) it easy to see that
 \[ \ker B_1|_{\la=\beta_1}=\im B_2|_{\la=\beta_2},\quad
  \ker B_2|_{\la=\beta_2}=\im B_3|_{\la=\beta_3},\ldots \]
Moreover, these matrices satisfy the consistency conditions (\ref{qmn}) (see
 Lemma below) and
   \[ B_2B_1= \la-\beta_1+\beta_{12}P_{12},\quad
 B_3B_2=
  \la-\beta_2+\beta_{23}P_{23},\dots,\quad \beta_{ij}\defeq\beta_i-\beta_j,
 \]
 where $P_{ij}^2=P_{ij}$ are projectors. In particular,
\[ \beta_{12}P_{12}=\left(\ba{cc}
 \frac14v_2u_{13} & \frac12 u_{13}\\
 \frac12v_2(\beta_{12}-\frac14v_2u_{13}) & \beta_{12}-\frac14v_2u_{13}
  \ea\right),\quad
u_{13}\defeq u_1-u_3,\]
\[ \beta_{23}P_{23}=\left(\ba{cc}
\beta_{23}+\frac14u_2v_{42} & \frac12u_3(\beta_{32}+\frac14u_2v_{24})\\
 \frac12v_{42} & \frac14 u_3v_{24}\\
  \ea\right),\quad
v_{ij}\defeq v_i-v_j.\]
Thus these matrices provide us with a factorization of elementary Darboux
transformations (\ref{sha}).  Observe that, for odd $i$, the product
$B_{j}B_i,$ where $j=i+1$, depends only on $v_j,\, u_{ij}=u_i-u_j$ for $i$ odd 
and on the difference $v_{i+2}-v_{i+1}$ and $u_{i+1}$ for $i$ even.

In comparison with the factorization defined by main Theorem, the
entries of matrices (\ref{odd}) and (\ref{even}) are expressed
{\bf directly} in terms
 of the potentials
\[ U_j=\left(\ba{cc}
 \la &  u_j\\
 v_j & -\la
\ea\right)=\la\sigma+q_j
\]
of the $2\times2$ Zakharov-Shabat spectral problem
\be\label{zs}
 \psi^1_x-\la\Psi^1=u\Psi^2,\quad\Psi^2_x+\la\Psi^2=v\Psi^1,\quad
 \mbox{or}\quad \Psi_x=U\Psi.
\ee Thus, the substitution of matrices (\ref{odd}) and
(\ref{even}) into the equations (\ref{bub}) yields the dressing
chain for the spectral problem (\ref{zs}): 
\be\label{jodd}
(D+2\beta_j)v_{j+1}+2v_j=\frac12u_jv_{j+1}^2,\quad
(D-2\beta_j)u_j-2u_{j+1}=-\frac12u_j^2v_{j+1}, \ee \be\label{jeven}
(D-2\beta_j)u_{j+1}-2u_j=-\frac12u_{j+1}^2v_j,\quad
(D+2\beta_j)v_j+2v_{j+1}=\frac12u_{j+1}v_j^2. \ee 
The first case
corresponds to (\ref{odd}) and the second one to (\ref{even}).

\paragraph{Remark.} In the case of the general Zakharov-Shabat spectral problem
(\ref{ZS}) one could consider the following Darboux transformations (\ref{bub})
\[ B_j=\la\sigma_j+ b_j ,\quad U_j=\la\sigma+q_j.\]
Collecting quadratic and linear terms in $\la$ in (\ref{bub}) one
finds
\[  [\sigma,\sigma_j]=0,\quad
\sigma_{j,x}=[\sigma,A_j]+q_{j+1}\sigma_j-\sigma_j q_j \]
That implies that $\sigma_{j,x}=0$ since the potentials $q_j$
are off-diagonal in the basis where the matrix $\sigma$ is diagonal.
Thus, the matrices $\sigma_j$ should be $x-$independent and could be any
diagonal matrix in that basis.
\vspace{2mm}

 Coming back to the spectral problem (\ref{zs}) we will show how,
starting with $(u_1,v_1)$, one construct a chain of Darboux
matrices $B_j,$ transforming the equation (\ref{zs}) with the
potential $U_j$ into one with the potential $U_{j+1},$ using the solution to the
corresponding spectral problem Riccati equation:
 \be\label{ricc}
 \rho_x=2\la\rho-v_j\rho^2+u_j,\quad \rho\defeq \frac{\Psi^1}{\Psi^2}.
\ee This should highlight a similarity with the dressing chain
considered in \cite{VeSh}.

\paragraph{Proposition.} There is a one-to-one correspondence between the
matrices (\ref{odd}), (\ref{even}) and solutions of the Riccati equation
(\ref{ricc}) with $\la=\beta_j.$

\noindent{\bf Proof:} At the first step we have to find the potentials $u_2$
and $v_2.$ The function $v_2$ is defined by (\ref{jodd}) as solution of ODE
\[ (D+2\beta_1)v_{2}+2v_1=\frac12u_1v_{2}^2 \]
which is reduced by rescaling $v_2=2\rho$ to (\ref{ricc}) with $\la=\beta_1.$
For the function $u_2$, we obtain now the exact formula
 \[2u_2=u_{1,x}-2\beta_1 u_1+\frac12 u_1^2v_2. \]

At the next step, we find $u_3=2\rho^{-1}$ by solving (\ref{ricc}) with
$\la=\beta_2.$ It satisfies the corresponding equation (\ref{jeven}):
\[  (D-2\beta_2)u_{3}-2u_2=\frac12u_3^2v_{2} \]
while $v_3$ is defined by the second equation in (\ref{jeven}). Thus in both
cases the system of equations (\ref{jodd}) and (\ref{jeven}) are
reduced to the Riccati equation (\ref{ricc}).\qed

\paragraph{Consistency problem.} We will ignore now the dependence
on the continuous independent variable $x$ and consider more
general (as comparised with (\ref{odd}), (\ref{even})) matrices
linear in $\la$ and satisfying conditions analogous to
(\ref{qmn}). Having applications in mind, we will denote one of the
matrices in (\ref{qmn}) as $A(n,\la)$ and
 consider the second one as a Darboux transformation for the discrete spectral
 problem defined by $A:$
 \be\label{AB}
 \Psi_{n+1}=A(n,\la)\Psi_n,\quad
A=\left(\ba{cc}\la+c & a\\ b & \alpha \ea\right),\quad
  \hat\Psi=B\Psi, \quad  \hat\Psi_{n+1}=\hat{A}(n,\la)\hat \Psi.\ee
If the spectral problem in (\ref{AB}) is a ``good'' one,
\footnote{The particular case of $A$ with $\alpha=0$ is related
with Toda chain.} then there should exist nontrivial Darboux
transformations with \be\label{nea} B=\left(\ba{cc}\eps & f\\ g &
h-\la \ea\right) \ee and the consistency equations $T_1\circ
T_2=T_2\circ T_1$, where
 $T_1\Psi=A\Psi,\,\, T_2\Psi=B\Psi$ (Cf. (\ref{qmn})) yield
\[
(\la+\hat c)\eps+g\hat a=\eps'(\la+c)+f'b,\quad
 (\la+\hat c)f+\hat a(h-\la)=\eps'a+f'\alpha
\]\[
 \eps\hat b+\hat{\alpha} g=(\la+c)g'+b (h'-\la),\quad
\hat{\alpha}(h-\la)\eps+f\hat b=\alpha(h'-\la)+g'a\]
 where $B'\defeq T_1(B)$ and $\hat A\defeq T_2(A).$ We see that
\[ g'= T_1(g)=b ,\quad \hat{a}= T_2(a)=f,\quad \eps'=\eps,\quad
\hat\alpha=\alpha
\]
and thus the consistency equations are reduced to
\be\label{***}
\left\{\ba{cc}
b(h'+c)=\eps\hat b+\alpha g , & \eps(\hat c-c)=f'g'-fg\\
\hat{a}(h+\hat c)=\eps a+\alpha f', &
\alpha(h'-h)=\hat{a}\hat b-ab \ea\right. .
\ee

It is easy to verify now
\paragraph{Lemma.} The matrices $B_j$ and $B_{j+1}$ defined by (\ref{odd}),
(\ref{even}) satisfy the consistency equations (\ref{***}) for any $j.$

Generally we find that (Cf. (\ref{qmn}))
\[ A\circ B=
\left(\ba{cc}\eps(\la-\delta+h^\dagger) & \hat{a}h^\dag \\
bh^\dag &\alpha(h^\dag-\la+\beta)\ea\right)
\]
where
\[ h^\dag=\hat c+h=c+h',\quad ab=\alpha(c+\beta),\quad fg=\eps(h-\delta),
\det A=\alpha(\la-\beta_1),\quad \det B=\eps(\beta_2-\la)
\]
and thus our Lemma corresponds to the case $\eps+\alpha=0.$

   The Darboux matrix $B$ in (\ref{AB}) is reciprocal to $A$ and (unlike the
   continuous case) one can replace it with $A\circ B.$ This property
   allows us to use the matrix $A$ as a starting point in the spectral problem (\ref{AB})
   with $\alpha=0$ \cite{99}.

\paragraph{An $R$-matrix interlude.}
While working out examples for the above factorization theorem, we
noticed that some of them were also solutions of the quantum
Yang-Baxter equation \be\label{YB}
R_{12}(\la-\mu)R_{13}(\la)R_{23}(\mu)=
R_{23}(\mu)R_{13}(\la)R_{12}(\la-\mu). \ee Here $R_{12}=R\otimes
Id,$ $R_{23}=Id \otimes R$ and $R_{13}=(Id \otimes P) R_{12} (Id
\otimes P)$ acts trivially in the second of the three spaces, $P$
is the usual permutation operator:
\[ P(u \otimes v)\defeq v \otimes u.  \]
Thus in the simplest $2\times2$ case, (\ref{YB}) is an equation for a
$4\times4$ matrix $R$ and a well known solution is given by
\be
R(\la)=\left(\ba{cccc}
 \la+i \eta & 0 & 0 & 0 \\
 0 & \la & i\eta & 0 \\
 0 & i\eta & \la & 0 \\
 0 & 0 & 0 & \la+i\eta
\ea\right)=\frac{1}{2} \left[ (\la+i \eta)P+(\la-i \eta)(1-P)
\right]. \label{ff}
\ee
and corresponds to the $XXX-$spin chain.
Since $P$ is a projector, we can express it in the form:
\begin{equation}
P=\frac{ |a \rangle \langle b |}{\langle a | b \rangle}, \qquad \mbox{where} \quad 2
\langle a|= \langle b| =(0,1,-1,0) \label{pp}
\end{equation}
We found that there exist two other possible choices of $\langle
a|$ and  $\langle b|$ in (\ref{pp}) such that an $R$-matrix of the
form
\begin{equation}
R(\la)=(\la+i \eta)P+(\la-i \eta)(1-P) \label{RP}
\end{equation}
satisfies the Yang-Baxter equation, namely:
\[ \langle a |=(\alpha^{-1},\, 1,\, 1,\, \alpha),\quad
 \langle b |=(\beta^{-1},\, 1,\, 1,\, \beta),
 \]
\[  \langle a |=(-1,\, \alpha,\, \beta,\, -\beta\alpha),\quad
\langle b |=(\beta\alpha,\, \alpha,\, \beta,\, 1),
 \]
with $\alpha, \beta$ being two real parameters.

Having two new solutions of the Yang-Baxter equation, we can
construct the associated Lax matrices and, finally, two monodromy
matrices out of them. Then, taking the traces of the monodromy
matrices we obtain two sets of commuting quantum observables.
Unfortunately the elements of each set are functionally dependent. 
Hence, these two new solutions do not give rise to new
quantum integrable systems.

Further investigations on Quantum Yang--Baxter equation showed
that, if we generalize the $R$--matrix form (\ref{RP}) dropping
the normalization condition (\ref{condition2}), then other
solutions, linear in the spectral parameter $\lambda$, can be found.
We notice that $R$ can then be expressed in the form $R=\la A +B$,
where $A$ and $B$ must be solutions to the two-dimensional
constant quantum Yang-Baxter equation; all such solutions have
been found by J. Hietarinta \cite{Hietarinta}.

An example of solution that we believe could be of some interest
is the following one:
\begin{displaymath}
R(\lambda)=\left(
\begin{array}{cccc}
\eta & 0 & 0 & \lambda \\
0 & 0 & \lambda+\eta & 0 \\
0& \lambda+\eta & 0 & 0 \\
\lambda & 0 & 0 & \eta
\end{array}
\right), \qquad R^{-1}=\frac{1}{\lambda^2-\eta^2}
\left(
\begin{array}{cccc}
-\eta & 0 & 0 & \lambda \\
0 & 0 & \lambda-\eta & 0 \\
0& \lambda-\eta & 0 & 0 \\
\lambda & 0 & 0 & -\eta
\end{array}
\right)
\end{displaymath}
Work is in progress on this topic.

\paragraph{Acknowledgements}
We would like to thank Decio Levi and Antonio Degasperis for
useful discussions and the interest shown in this paper.

This work was partially supported by Russian Foundation for Basic
Research, grants No. 04-01-00403 and Nsh 1716.2003.1.

 \end{document}